\documentclass[twocolumn]{aa} 
\usepackage{natbib}
\usepackage{amsmath,amssymb,amsfonts,bm}
\usepackage{subfigure}
\usepackage{longtable}
\usepackage{graphicx}

\begin{document}
\title{Spin-up of massive classical bulges during secular evolution}

\author{Kanak Saha \inst{1,2}, Ortwin Gerhard \inst{2} \and Inma Martinez-Valpuesta \inst{3}}

\institute{
	Inter-University Centre for Astronomy and Astrophysics, Post Bag 4, Ganeshkhind, Pune 411007, India\\ \email{kanak@iucaa.in}, 
	\and
	Max-Planck-Institut f\"{u}r Extraterrestrische Physik, Giessenbachstrasse, 85748 Garching, Germany\\ 
        \email{gerhard@mpe.mpg.de}
        \and 
        Instituto de Astrofisica de Canarias, E-38205 La Laguna, Tenerife, Spain\\
        \email{imv@iac.es} 
           }

\authorrunning{Saha et al.}
\titlerunning{Spin up of massive classical bulges}

\abstract
{Classical bulges in spiral galaxies are known to rotate but the
  origin of this observed rotational motion is not well understood. It
  has been shown recently that a low-mass classical bulge (ClB) in a
  barred galaxy can acquire rotation from absorbing a significant
  fraction of the angular momentum emitted by the bar.}
{Our aim here is to investigate whether bars can spin up also more
  massive ClBs during the secular evolution of the bar, and to study
  the kinematics and dynamics of these ClBs.}
{We use a set of self-consistent N-body simulations to study the
  interaction of ClBs with a bar that forms self-consistently in the
  disk. We use orbital spectral analysis to investigate the angular
  momentum gain by the classical bulge stars.}
{We show that the ClBs gain, on average, about 2 - 6\% of the disk's
  initial angular momentum within the bar region.  Most of this angular
  momentum gain occurs via low-order resonances, particularly $5:2$
  resonant orbits. A density wake forms in the ClB which corotates and
  aligns with the bar at the end of the evolution. The spin-up process
  creates a characteristic linear rotation profile and mild tangential
  anisotropy in the ClB.  The induced rotation is small in the centre
  but significant beyond $\sim2$ bulge half mass radii, where it leads
  to mass-weighted $V/\sigma \sim 0.2$, and reaches a local $V_{\rm
    max}/\sigma_{\mathrm{in}}\sim0.5$ at around the scale of the
  bar. The resulting $V/\sigma$ is tightly correlated with the ratio
  of the bulge size to the bar size.  In all models, a box/peanut
  bulge forms suggesting that composite bulges may be common.}
{Bar-bulge resonant interaction in barred galaxies can provide some
  spin up of massive ClBs, but the process appears to be less
  efficient than for low-mass ClBs. Further angular momentum transfer
  due to nuclear bars or gas inflow would be required to explain the
  observed rotation if it is not primordial.}


\maketitle

\section{Introduction}
\label{sec:introduc}
Classical bulges (hereafter ClBs) are the central building blocks in
many early-type spiral galaxies. ClBs might have formed as a result of
major mergers during the early phase of cosmic evolution
\citep{Kauffmanetal1993, Baughetal1996, Hopkinsetal2009, Naabetal2014}, or through a
number of other mechanisms such as monolithic collapse of primordial
gas clouds \citep{Eggenetal1962}, the coalescence of giant clumps in
gas-rich primordial galaxies \citep{Noguchi1999,Immelietal2004,
  Elmegreenetal2008}, violent disk instability at high-redshift \citep{Ceverinoetal2015}, 
multiple minor mergers \citep{Bournaudetal2007, Hopkinsetal2010}, 
and accretion of small companions or satellites
\citep{Aguerrietal2001}. Although most of these studies do not provide
quantitative predictions for the bulge kinematics, it is generally
believed that ClBs formed through these processes have low rotation
compared to the random motion. 
\noindent For example, \cite{Naabetal2014} showed that spheroids produced 
by minor and major mergers (which include ClBs) in full cosmological hydrodynamical
simulations have a wide range of rotational properties, with the massive ones
having $V/\sigma$ less than 0.5. \cite{Elmegreenetal2008} reported
dispersion dominated clump-origin ClBs with upper limit on $V/\sigma
\sim 0.4 - 0.5$, where $V$ is the rotation velocity and $\sigma$ is
the central velocity dispersion. A similar study by \cite{InoueSaitoh2012} 
suggests that clump-origin bulges have exponential like surface density profiles and 
rotate rapidly with $V/\sigma \sim 0.9$, resembling pseudobulges 
\citep{KormendyKennicut2004}. However, using cosmological hydrodynamical simulations 
with continuous gas accretion, \cite{Ceverinoetal2015} showed that  
massive classical-like bulges with non-zero angular momenta are produced at high redshift 
but provided no estimate on the bulge $V/\sigma$. Overall, there is a lack of clear quantitative 
picture of the rotational motion induced during the formation of classical bulges in 
numerical simulations. 

Various observational measurements have confirmed that ClBs in spiral
galaxies possess rotation about their minor axis and in most cases in
the same sense as the disk rotates \citep{KI1982, Cappellarietal2007,
  Fabriciusetal2012}. Disk galaxies both barred (e.g., NGC 1023, NGC
3992) and unbarred (e.g., NGC 4772, NGC 2841) host ClBs with a wide
range of masses and sizes \citep{KI1982,
  K1982,Laurikainenetal2007,Cappellarietal2007}. More recently,
\cite{Fabriciusetal2012} have obtained detailed kinematic observations
of a large sample of bulges including ClBs (whose classification is
based primarily on the bulge Sersic index).  From these studies it is
known that ClBs rotate typically with $V/\sigma$ close to an oblate
isotropic rotator model \citep{Binney1978}, i.e., faster than
low-luminosity elliptical galaxies, but more slowly than
pseudobulges. The origin of such rotational motion observed in ClBs
remains unclear.

Recent cosmological hydrodynamical simulations which include feedback
and smooth accretion of cold gas through cosmic filaments show that
exponential disks could have assembled around merger-built ClBs and
grown through the galaxy's assembly history
\citep{Governatoetal2007,Agertzetal2011, Brooketal2012}. Once such a
disk becomes massive and dynamically cold, its self-gravity may induce
a bar instability leading to rapid formation of a bar which would then
interact with the preexisting ClB \citep{HernquistWeinberg1992,
  Athanamisi2002,Sahaetal2012}. In particular, \cite{Sahaetal2012}
showed that an initially non-rotating low-mass ClB could absorb a
significant fraction of the angular momentum emitted by the bar and
end up rotating within $\sim2$ Gyr. However, it is not clear yet how
more massive ClBs would react to the bar.

The goal of this paper is to investigate whether angular momentum
transfer from the bar is an important process for the origin of
angular momentum in ClBs embedded in spiral galaxies. For this, we use
collisionless N-body simulations to study the secular evolution in a
set of model galaxies with initially non-rotating ClBs with different
masses and sizes. Our simulation results show that, indeed, massive
ClBs could be spun up in their outer parts, typically in less than half
of the Hubble time. In these models, a box/peanut bulge always forms
alongside the secular evolution, suggesting that composite bulges
may be common.

The paper is organized as follows. In Section~\ref{sec:modelsetup}, we
describe the initial galaxy models and the set-up for the {\it N}-body
simulations.  Section~\ref{sec:growth} describes the growth of a bar
and its size evolution in these models. The transfer of angular
momentum and the orbital analysis to account for the resonant trapping
are described in Section~\ref{sec:angularMom}. The kinematic
properties of the these bulges are described in Section~\ref{sec:kinematics}. 
Finally, a discussion of the results and the conclusions from this work 
are presented in Section~\ref{sec:discus}.

\section{Initial galaxy models with massive bulges}
\label{sec:modelsetup}
We present here a set of $7$ galaxy models constructed
using the self-consistent method of \citet{KD1995}. Each galaxy model
consists of a live disk, dark matter halo and bulge. The initial disk
is modelled with an exponentially declining surface density in the
radial direction with a scale-length $R_d$, mass $M_d$ and with a
${\rm sech}^2$ distribution of stars with vertical scale-height $h_z$. The initial
radial velocity dispersion of the disk stars follows an exponential
distribution with a radial scale length $R_{\sigma}=0.5 R_d$ such that
the disk has a constant scale height throughout. The live dark matter
halo is modelled with a lowered Evans model and the classical bulge
with a King model. The dark matter halo has a core in the inner
regions of the galaxy model and produces a nearly flat rotation curve
in the outer parts of the disk. For details on constructing these
models, the readers are referred to \citet{Sahaetal2010, Sahaetal2012}.

\begin{figure}
\rotatebox{0}{\includegraphics[height=7.0 cm]{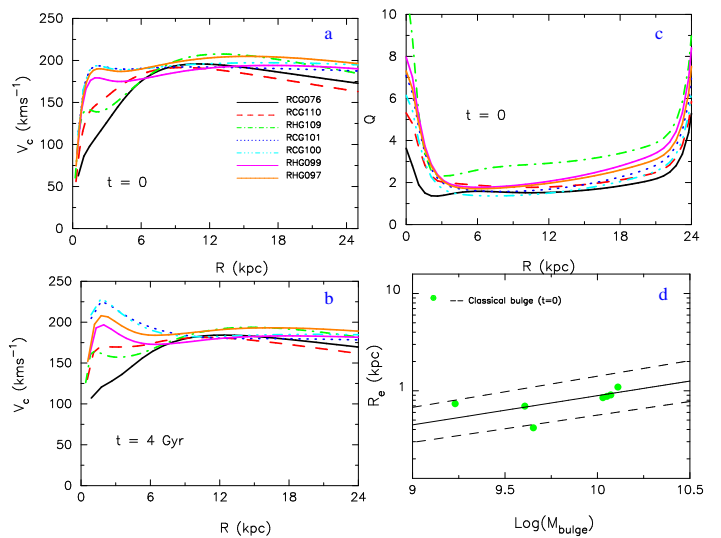}}
\caption{Initial total circular velocity curves (a), Toomre Q profiles (c), 
the total circular velocity curves at $4$~Gyr (b) and Mass-size relation for
the initial ClBs (d).}
\label{fig:vcQt0}
\end{figure}

We scale each model such that the initial disk scale length $R_d =
4$~kpc and the circular velocity at $\sim 8.5$~kpc is $\sim
200$~kms$^{-1}$. The initial circular velocity curves for all the
models are shown in Fig.~\ref{fig:vcQt0}(a). Each model has a characteristic 
profile for the Toomre $Q(R) = \sigma_r(R)\kappa(R)/{3.36 G \Sigma(R)}$, where
$\sigma_r$, $\kappa$ and $\Sigma$ denote the radial velocity dispersion, 
epicyclic frequency and the surface density of stars; see Fig.~\ref{fig:vcQt0}(c). 
The stellar disks in our sample range from dynamically cold to fairly hot. 
Table~\ref{tab:paratab} summarizes the initial parameters of the disk, bulge and dark halo.

The bulges in our galaxy models are initially non-rotating and
dispersion dominated. The Sersic indices for these King model bulges
are generally close to $1$ or less. Recent findings by show that barred 
galaxies can have classical bulge like components with sersic indices close 
to 1 or less \cite{Laurikainenetal2007,Erwinetal2015}. The density 
profiles in the inner region are flatter than $R^{{1/4}}$ profiles which 
are known to represent the surface brightness profiles of traditional ClBs
\citep{deVaucouleurs1953, FisherDrory2010}. Unlike $R^{{1/4}}$ profiles, the
density profile of a King model bulge has a well-defined outer boundary, 
defined by a truncation radius $R_b$ given in table~\ref{tab:paratab} and
other well-studied properties which motivated us to use them as a
simple model for the bulges. Setting up of the stellar kinematics for the bulge
stars is relatively straightforward since it is modelled by an analytic DF 
(King model, $f_{b}(E)$) that depends only on the energy integral ($E$). 
Assigning a velocity to a bulge star is done in the following way: corresponding to 
a star's position, we find the local maximum of the DF and then employ acceptance-rejection 
technique to find a velocity. The 3 components of a star's velocity ($v_x, v_y, v_z$) are randomly 
selected from a velocity sphere with radius equal to the local escape velocity. The velocity 
ellipsoid is isotropic by construction. The DF accepts 3 free parameters of which, $\sigma_b$
determines the velocity dispersion of the bulge stars; higher the $\sigma_b$ larger the value
of the velocity dispersion. Modelled in this way, our initial ClBs are kinematically hot
spheroidal stellar systems that are similar to typical observed ClBs. 

To further compare our initial ClBs with observed ClBs, we performed a 2D 
bulge-disk decomposition of our initial galaxy models. We first made FITS files 
for all the galaxy models and analysed them with GALFIT \citep{Pengetal2002}. In
Fig.~\ref{fig:vcQt0}(d) we show the effective radii and masses for our
initial ClBs, and compare them with the relation $R_e \propto
M_b^{\alpha}$ with $\alpha =0.3$ for observed ClBs found by
\citet{Gadotti2009}. Most of our initial ClBs follow this relation
closely, while models RHG109 and RCG076 are $1 \sigma$ away from the
mean relation.

\begin{table}
  \caption[ ]{Initial bulge, disk and halo model parameters, ordered by the 
    ratio of ClB to disk mass. Column (1): model name. (2): Ratio of
    bulge to disk mass, $M_b/M_d$. (3), (4): tidal radius $R_b$ and half-mass 
    radius $R_{{b,1/2}}$ of the initial ClB, normalized by disk scale-length $R_d$. 
    (5): $Q$ parameter (at $2.5 R_d$). (6): disk mass. (7): mass of dark matter halo $M_h$, 
    in units of $M_d$. (8): contribution
    of bulge and disk together to the total circular velocity at $2.2 R_d$.}
\begin{center}
\begin{tabular}{lccccccccccccccc}  \hline\hline
  Models & $\frac{M_b}{M_d}$ & $\frac{R_b}{R_d}$ & $\frac{R_{{b,1/2}}}{R_d}$ & $Q$ & $M_d$ & $\frac{M_h}{M_d}$  &  $\frac{V_{{c,bd}}}{V_{{c,tot}}}$  \\
           & &   &  & & ($10^{10} M_{\odot}$) & \\
\hline
\hline
RCG076 & 0.11 & 2.1  & 0.2    &1.5 & 2.0 & 11.0 & 0.5 \\
RCG110 & 0.16  & 0.8  & 0.2  &1.8 & 2.7 & 5.5  & 0.6  \\
RHG109 & 0.18  & 0.6  & 0.14  &2.8 & 2.7 & 6.9  & 0.6  \\
RCG101 & 0.23  & 1.1  & 0.2  &1.6 & 5.1 & 6.7  & 0.8 \\
RCG100 & 0.27  & 1.1  & 0.2   &1.4 & 4.6 & 7.5  & 0.8 \\
RHG099 & 0.39  & 1.1  & 0.2  &1.9 & 3.2 & 10.0  & 0.7 \\
RHG097 & 0.43 & 1.2  & 0.2   &1.8 & 3.1 & 7.9 & 0.7 \\
\hline
\end{tabular}
\end{center}
\label{tab:paratab}
\end{table}

\begin{figure}
\rotatebox{0}{\includegraphics[height=9.3 cm]{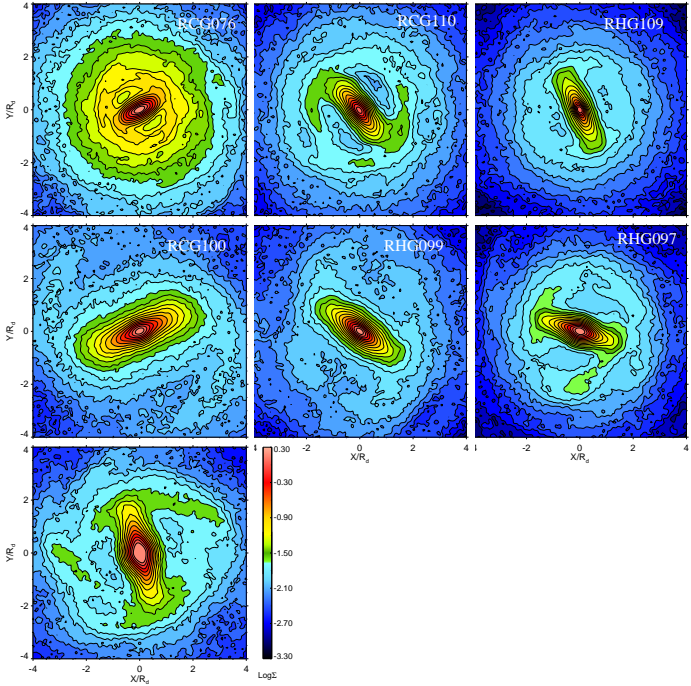}}
\caption{Face-on surface density maps for all model galaxies after $4$~Gyr of
  evolution.}
\label{fig:surfden}
\end{figure}

We evolve each galaxy model in isolation to examine the evolution of
the bulge shape, morphology and kinematics. The simulations are
performed using the Gadget code \citep{Springeletal2001} which uses a
variant of the leapfrog method for the time integration. The forces
between the particles are calculated using the Barnes \& Hut (BH) tree
with some modification \citep{Springeletal2001} with a tolerance
parameter $\theta_{tol} = 0.7$. We compute the virial 
equilibrium condition of a galaxy model at different times using 
the Gadget output and found that some of the models were although 
initially out of equilibrium (by about a few percent) settles down
close to virial equilibrium (by about a percent or less) within a
rotational time scale.  
A total of $2.2 \times 10^6$ particles
is used to simulate each model galaxy of which $1.0 \times 10^5$ are
in the ClB, $1.05 \times 10^6$ in the disk and $1.05 \times 10^6$ in
the dark matter halo. The softening lengths for the disk, bulge and
halo are all unequal and are chosen so that the maximum force from
particles of all species (bulge, disk, halo) is nearly the same
\citep{McMillan2007}.

\begin{figure}
\begin{center}
\rotatebox{270}{\includegraphics[height=7.0 cm]{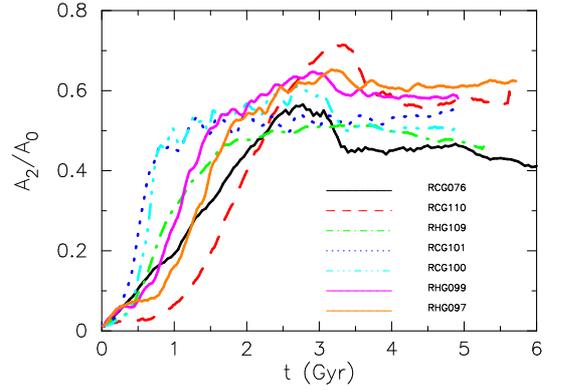}}
\caption{Time evolution of the bar strength in our galaxy models with
  pre-existing classical bulges.}
\label{fig:A2}
\end{center}
\end{figure}

\section{Growth of a bar and its size evolution}
\label{sec:growth}
Although about $70$\% of disk galaxies in the local universe are barred
\citep{Eskridgeetal2000,Barazzaetal2008}, the formation of a bar is still 
not fully understood. The growth rate of a bar in a disk galaxy depends on 
various parameters of the disk and its gravitational interaction with surrounding
the dark matter halo and preexisting ClB (if present). Numerous studies of N-body 
simulations show
that a bar forms and grows rapidly in a cold rotating stellar disk
with Toomre $Q$ close to $1$ \citep[and references
therein]{Hohl1971,SellwoodWilkinson1993,Athanassoula2002,Dubinskietal2009}.
Swing amplification \citep{Toomre1981} and cooperation of orbital
streams \citep{Earn-LyndenBell1996} are thought to play a key role
in the bar growth. The linear growth of a bar is directly affected by the Toomre $Q$ of the
stellar disk as can be seen from Fig.~\ref{fig:A2}. A bar grows rapidly in a cool disk, 
whereas it grows rather slowly in a hotter disk \citep{Sahaetal2010}. The slow growth 
of a bar occurs through non-linear processes; in particular, the resonant gravitational
interaction with surrounding dark matter halo is known to play a major role. 
This has been investigated by several authors in the past \citep{DebattistaSellwood1998,
Athanassoula2002,Holley-Bockelmannetal2005,WeinbergKatz2007a,Ceverinoklypin2007,SahaNaab2013b}.

\begin{figure*}
\begin{center}
\rotatebox{0}{\includegraphics[height=6.5 cm]{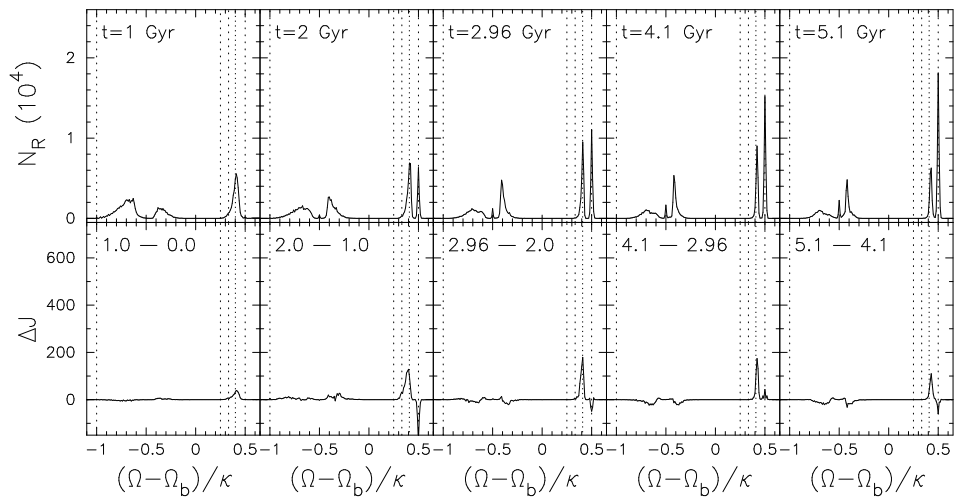}}
\caption{The upper panels show the distribution of bulge stars with
  frequency $(\Omega-\Omega_B)/\kappa$ at different times throughout
  the secular evolution in model RHG097. The lower panels show the net
  change in the angular momentum of the selected stars with respect to
  the previous time.  The vertical dotted lines indicate the most
  important resonances: from left they are $-1:1$, $4:1$, $3:1$, 
    $5:2$ and $2:1$. As time progresses, more stars are trapped
  by the ILR of the bar in the stellar disk. However, most of the
  angular momentum transfer occur through the $5:2$ resonance.}
\label{fig:reso97}
\end{center}
\end{figure*}

The role of a pre-existing ClB alone on the formation and evolution of
a bar has not been fully investigated. Bar formation could, in
principle, be hampered by a central mass concentration
\citep{Hasanetal1993,SellwoodMoore1999,Athanassoulaetal2005}, because
the bulge could cut the feedback loop required for swing amplification
by placing an ILR (inner Lindblad resonance) near the center of a
galaxy.  In previous studies \cite[e.g.,][]{Sahaetal2012}, a low mass
initially non-rotating ClB facilitated the bar growth by absorbing
angular momentum through the ILR.  Later, \cite{SahaGerhard2013a}
investigated the effect of varying degree of initial rotation in the
preexisting ClB. But a detailed investigation of the effect of massive
bulges on bar formation is yet to be done.

All the galaxy models in our simulation sample have a preexisting ClB
of different mass and size (see Table~\ref{tab:paratab}), and they all
form a bar in their stellar disks. Fig.~\ref{fig:surfden} shows
surface density maps for all stars in each galaxy model at the end of
$4$~Gyr. The models are arranged such that the initial value of the
bulge-to-disk mass ratio increases gradually from the top left corner
(RCG076) of Fig.~\ref{fig:surfden} to the bottom right corner
(RHG097). Note that Toomre Q alone is not the factor deciding the final 
outcome of these models. Dark matter distribution and the bulge might also 
play an important role. Most of these models do not form long-lived two-armed 
spirals.

In Fig.~\ref{fig:A2}, we show the time evolution of the bar amplitude
($A_2/A_0$) in all our models. $A_2$ and $A_0$ are the $m=2$ and
$m=0$ Fourier component of the surface density respectively. The
fastest bar growth is seen in model RCG100 (Q=1.4). From Fig.~\ref{fig:vcQt0}(a)
and Fig.~\ref{fig:A2}, we see that generally the steeper the slope of
the initial rotation curve, the faster the growth rate of a bar (see
model RCG100 versus RCG099 and RHG076). However, the correspondence
between growth rate and slope of the rotation curve is not one-to-one.
For instance, the linear growth rate of the bar in RCG076 is higher than
RCG110 although the latter has a steeply rising rotation curve.
Faster bar growth is also supported by higher values of
$\frac{V_{{c,bd}}}{V_{{c,tot}}}$ (see Table~\ref{tab:paratab}). Note
the bar in model RHG109 (Q=2.8) is weakest of all.  

Once formed, the size of a bar increases as it exchanges angular
momentum with the surrounding dark matter halo and the pre-existing
bulge \citep{Sahaetal2012}.  In our simulations, we calculate the bar
sizes at different epochs using the method explained in \cite{SahaGerhard2013a}.
As these bars grow and become self-gravitating, they undergo the well
known buckling instability which leads to the formation of a
boxy/peanut bulge \citep{CombesSanders1981,PfennigerNorman1990, Rahaetal1991, MV2004}.  
The galaxy models with prominent boxy/peanut
bulges end up with a peak in the rotation curve in the boxy bulge
region, see Fig.~\ref{fig:vcQt0}(b).  The rotation velocities shown in
Fig.~\ref{fig:vcQt0}(b) are calculated at the end of $4$~Gyr; by that time
nearly all the bars are fully grown and have gone through the buckling
instability. The final bulges in our simulations are thus composite bulges 
which are a superposition of a ClB and a boxy/peanut bulge formed from the
disk stars. 

\begin{figure*}
\begin{center}
\rotatebox{0}{\includegraphics[height=6.5 cm]{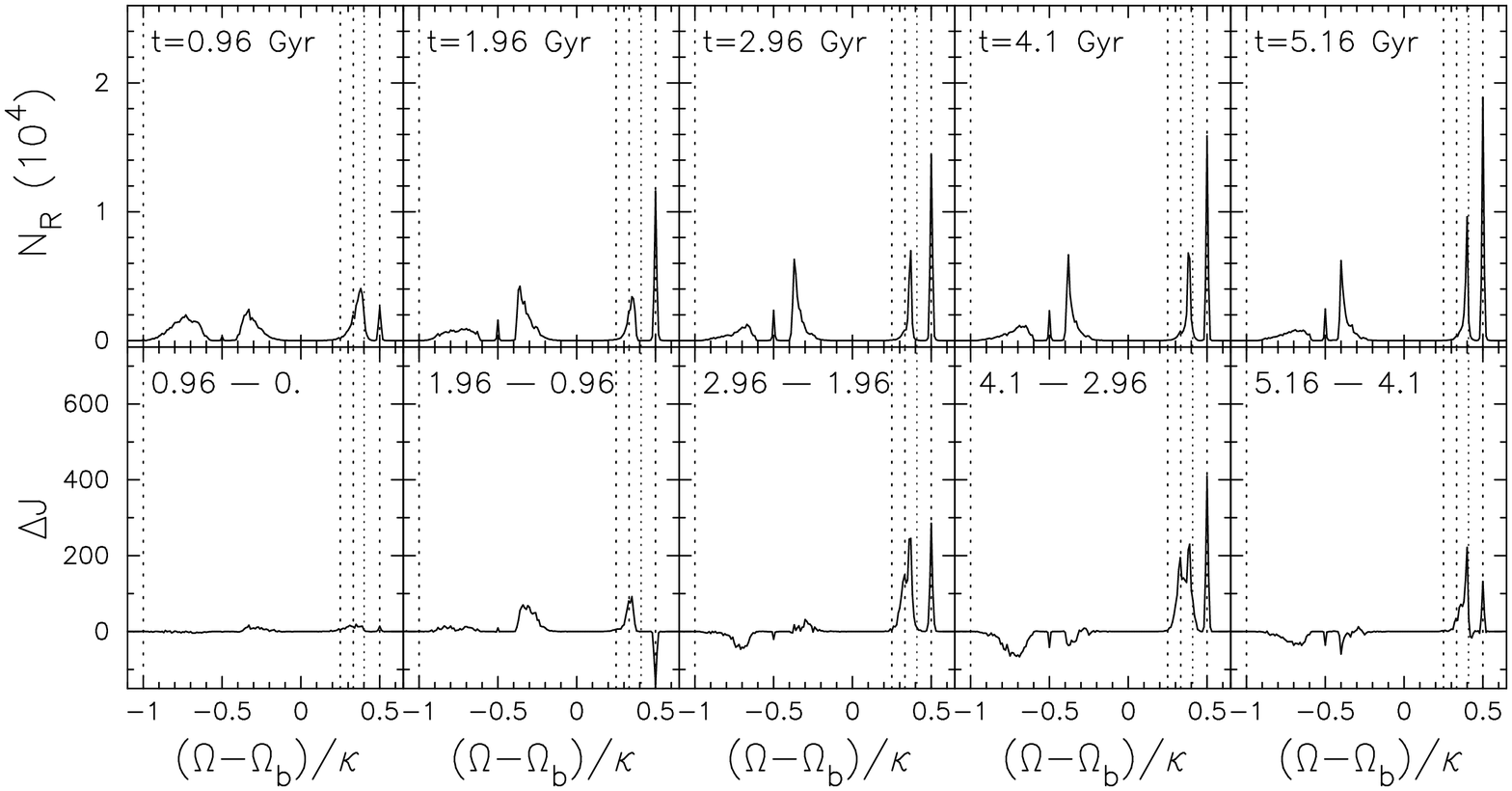}}
\caption{Same as in Fig.~\ref{fig:reso97} but for model RCG076.}
\label{fig:reso76}
\end{center}
\end{figure*}

\section{Angular momentum transfer and orbital analysis}
\label{sec:angularMom}
A bar grows by trapping more and more orbits in its main orbital
families (e.g., x1 orbit family). While it does so, the stars loose
angular momentum to the dark matter halo and the ClB in the model.
Since the distribution function, $f(E)$, of the initial ClB is described
by a King model where it is a function of energy alone, the gain of
angular momentum by the bulge at a given resonance is always positive
\citep{Sahaetal2012}. The dark matter haloes in our model galaxies
also gain angular momentum but here we are concerned with
investigating the structural and kinematic changes in the ClB that are
brought about by the angular momentum gain.
To understand the impact of a bar on the pre-existing ClB, we 
compute the angular momentum transfer between the bar and the bulge in our 
simulation using orbital spectral analysis \citep{BinneySpergel1982,
MV2004,Martinez-Valpuestaetal2006}. We focus, in particular, on two models RHG097 
and RCG076 which have the most massive and the least massive ClBs in our sample, 
with $M_b/M_d =0.43$ and $0.11$ respectively (see Table~\ref{tab:paratab}). 

Fig.~\ref{fig:reso97} shows the results of the orbital spectral
analysis for the bulge stars in model RHG097 at different times during
the evolution. As in the case of a low-mass ClB, \cite[see][]{Sahaetal2012},
more and more bulge stars get trapped by the $2:1$ resonance of the
rotating potential. However, unlike the low-mass case, not much
angular momentum is gained through the ILR of the bar by this ClB. In
fact, at $\sim 2$~Gyr (2nd panel in Fig.~\ref{fig:reso97}), the bulge
stars are seen to be loosing angular momentum through the ILR and this
continues subsequently with gradually diminishing magnitude (except at
$\sim 4$~Gyr, fourth panel from left). In essence, we find that the
massive ClB in RHG097, on average, loses angular momentum through the
2:1 resonance.  However, and interestingly, we find that a lot of
angular momentum is gained via 5:2 resonant orbit families
(corresponding to the peak around $(\Omega-\Omega_b)/\kappa =0.4$ in
Fig.~\ref{fig:reso97}), and this occurs consistently throughout the
evolution. In N-body bars, these $5:2$ orbits are known to be stable
periodic orbits like the $3:1$ and $4:1$ families inside the
corotation of the bar \citep{Voglisetal2007}.

\begin{figure}
\begin{center}
\rotatebox{0}{\includegraphics[height=8.0 cm]{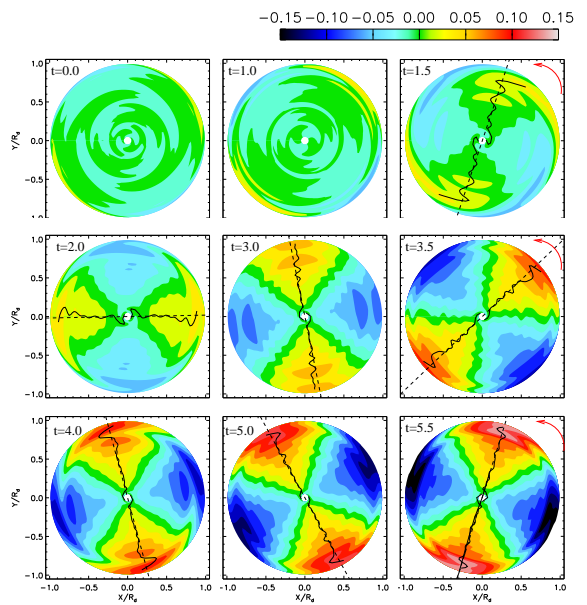}}
\caption{The $l=2, m=2$ density wake mode in the ClB of RHG097.
  Dashed lines in all panels indicate the location of the bar in the
  stellar disk while solid lines denote the phase angle corresponding
  to the 'bar' mode in the classical bulge (ClB). Towards the later
  phases of the evolution, the ClBb becomes progressively
  stronger. The red arrow denotes the rotation of the disk bar. Note that at 
1 Gyr, a clear bar has not developed yet. The unit of time is in Gyr.}
\label{fig:l2m297}
\end{center}
\end{figure}

Fig.~\ref{fig:reso76} shows the results from the spectral analysis of
model RCG076. We found a similar trend as in model RHG097 -- a
fraction of ClB stars in both models is gradually trapped at the $2:1$
resonance of the bar. This persists in all ClBs in our simulations
(see, Table~\ref{tab:paratab}). However, the details and the dominant
mode of angular momentum transfer vary. At $t=1.96$~Gyr (second panel
from left of Fig.~\ref{fig:reso76}), the initial bulge of RCG076 loses
angular momentum through the $2:1$ resonance. But at subsequent times,
it gains angular momentum consistently through $2:1$ orbits. This ClB
also gains angular momentum through other resonances: during the early
phase of bar growth, it gains angular momentum through the $3:1$
resonant orbits but gradually, the peak shifts from $3:1$ to $5:2$
resonance (see the right-most panel in Fig.~\ref{fig:reso76}). It is
interesting to notice that these resonances are rather broad compared
to the usual $2:1$ resonance - a feature that is common in both the
bulges of RHG097 and RCG076 as well as that in \cite{Sahaetal2012}.
In other words, associated with these resonances (e.g., $3:1$, $5:2$),
there are many off-resonant particles that contribute to the net
angular momentum transfer. While the two ClBs, in models RHG097 and
RCG076, gain angular momentum through $5:2$ resonance, we do not
see any angular momentum transfer through the $-1:1$ orbital families
outside corotation which contributed in the case of the low-mass ClB
studied by \cite{Sahaetal2012} using $10$~million particles. It is not clear to us whether
this is due to comparatively lower particle resolution in the current models 
($2.2$~million particles). Since low resolution noisy simulation could artificially
enhance the star-star encounters and knock stars off their resonant orbits,
\cite{Holley-Bockelmannetal2005, Dubinskietal2009} based on their explicit study 
on the model resolution and resonant behaviour, suggested that models with a 
few million particles generally show convergent behaviour.

\begin{figure}[b]
\begin{center}
\rotatebox{0}{\includegraphics[height=8.0 cm]{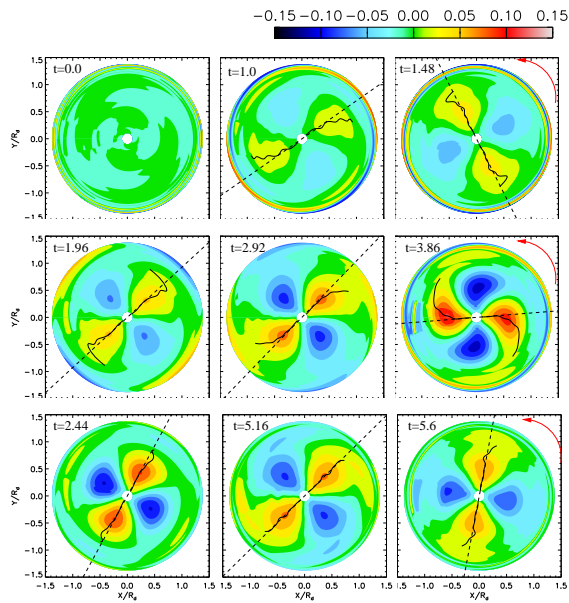}}
\caption{Same as in Fig.~\ref{fig:l2m297} but for RCG076.}
\label{fig:l2m276}
\end{center}
\end{figure}

Other than resonant orbit families, there is also a continuous
(chaotic orbits) component of angular momentum transfer of comparable
importance in both models RHG097 and RCG076 (around
$\Omega-\Omega_p/\kappa\simeq-0.3$ in Figs.\ref{fig:reso97} and
\ref{fig:reso76}). We have changed the numerical resolution to 
calculate the potential on a finer grid and repeated the above analysis 
to rule out numerical artifacts. Also \citet{Sahaetal2012}
found significant contributions to the net angular momentum gain of
the ClB via non-resonant chaotic orbits. For the two bulges in RHG097
and RCG076, the sign of the angular momentum transfer changes during
the evolution. They both gain angular momentum initially via the
non-resonant orbits, but in the later phases these bulges mostly loose
angular momentum via the stochastic orbits. Over the entire period of
evolution, the non-resonant stochastic orbits cause a net loss of
angular momentum from these bulge. However, we have verified that the
total angular momentum gain through resonant and non-resonant
orbits was positive at all times. In the following, we look at the 
density wakes created in these bulges by the bar.

\subsection{Density wakes and bulge harmonics}
\label{sec:wakes}
In this section, we examine the density wakes in the ClBs, primarily
the $l=2, m=2$ spherical harmonic modes, as the bulge stars interact
with the bar. Fig.~\ref{fig:l2m297} shows the density perturbation
corresponding to the $l=2,m=2$ spherical harmonic mode projected onto
the equatorial plane of the bulge at different times during the
evolution of RHG097. Such a density wake arises naturally as the bulge stars
interact gravitationally with the rotating bar potential. If dark matter
particles interact with the stellar bar, a so called 'halo-bar' gets created 
in the dark matter halo \citep{DebattistaSellwood2000, Holley-Bockelmannetal2005, 
SellwoodDebattista2006,Athanassoula2007, WeinbergKatz2007a,SahaNaab2013b}.
In the present case, we see that the perturbed density in the ClB corresponding to
this $l=2,m=2$ mode becomes stronger as time progresses, in compliance
with the increased number of trapped bulge stars at the $2:1$ resonance 
(upper panels of Fig.~\ref{fig:reso97} and Fig.~\ref{fig:reso76}). The disk bar 
rotates anticlockwise and its instantaneous position angle is denoted by the 
dashed line (see Fig.~\ref{fig:l2m297}). 

Initially, the density wake appears in the outer parts of the ClB of RHG097 
in the form of a spiralish feature (at about 1.5 Gyr when the disk bar forms) 
and become stronger eventually taking the shape of a bar-like mode, although 
not as clear as in the model RCG076 (shown below).
The phase of the $l=2,m=2$ mode in the bulge of RHG097 reveals that the disk-bar 
and the bulge-bar are nearly aligned with each other during
the entire evolution. Although there are instances when the two bars
are clearly misaligned by a very small angle (mostly in the outer parts, 
see Fig.~\ref{fig:l2m297}). Such a misalignment can cause dynamical
friction \citep{Chandra1943,TremaineWeinberg1984}, thereby leading to
the transfer of angular momentum either from the disk-bar to the bulge
or vice-versa. Fig.~\ref{fig:reso97} shows that there is little angular
momentum exchange via the $2:1$ resonance probably because of the small angle
misalignment between the two bars of RHG097.   

A similar analysis as above is shown in Fig.~\ref{fig:l2m276} for the
low-mass bulge model in RCG076. The development of a bulge-bar through
the $l=2,m=2$ mode follows a similar evolution as in model RHG097,
albeit with some differences. As time progresses, the bulge in RCG076
grows a comparatively more pronounced bulge-bar. Initially,
the density wake lags the disk-bar. At $t=1.48$~Gyr, a part of the 
bulge-bar appears to be leading the disk-bar. However, during the later 
phases of evolution, this bulge-bar lags behind the disk-bar and gains 
angular momentum through the $l=2,m=2$ mode, in compliance with the
angular momentum gain via $2:1$ resonant orbits (Fig.~\ref{fig:reso76}).
So in this case, we have a consistent picture of angular momentum transfer 
via the $2:1$ resonance and the misalignment of bulge-bar and disk-bar.

Inspection of Fig.~\ref{fig:l2m297} and Fig.~\ref{fig:l2m276} shows
that the peak of the $l=2,m=2$ density wake develops at different
spatial locations in the two bulges -- for the massive ClB (RHG097)
it is pronounced in the outer parts while for the low-mass case
(RCG076) it is confined to an intermediate radial range. Thus, only
the stars in the outer parts of the ClB in model RHG097 take part in
the rotational motion; see the line-of-sight velocity maps of
Fig.~\ref{fig:2Dmap97}. It is likely that both figure rotation and
streaming motion within the bulge-bar contribute to the net rotational
motion.

\section{Bulge kinematics}
\label{sec:kinematics}
The orbital configuration of the ClBs in our model galaxies changes as
a result of the angular momentum gain via the bar-bulge interaction,
and this must manifest itself in the kinematic properties of the ClBs.
In this section, we focus on the stellar kinematics, in particular on
the rotational motion of the classical bulge component induced by the
angular momentum gain. By separating the ClB particles from the
boxy/peanut bulge that forms from the disk through the bar buckling
instability, we investigate the dependence of the acquired rotational
motion in the ClBs on various parameters.
   
\subsection{Rotation and dispersion profiles}

\begin{figure}
\begin{center}
\rotatebox{0}{\includegraphics[height=7.0 cm]{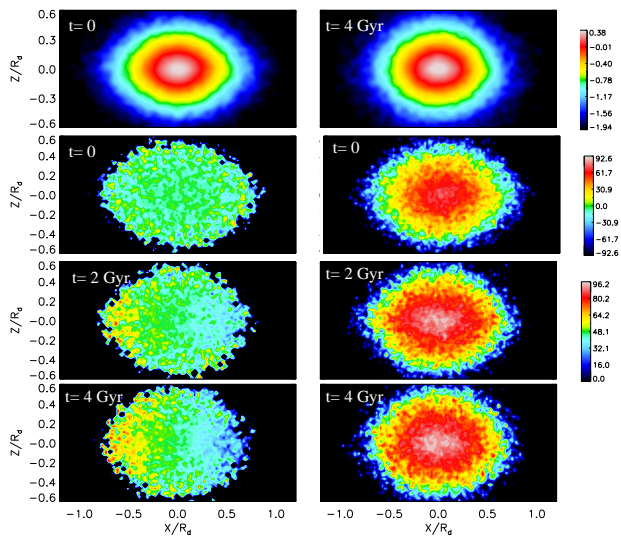}}
\caption{Surface density (upper panel), line-of-sight velocity and velocity
  dispersion maps (2nd to 4th panel) of the ClB in RHG097 at different times during the
  evolution. These images are taken at $90^{\circ}$ projection
  (edge-on view) and the major axis of the bar is aligned with the
  X-axis. Clear signatures of rotation are seen at $4$~Gyr. Color bar at top represents density,
middle the velocity and bottom the velocity dispersion.}
\label{fig:2Dmap97}
\end{center}
\end{figure}

In Fig.~\ref{fig:2Dmap97}, we show the evolution of the massive ClB
model RHG097 through surface density, line-of-sight velocity and
dispersion maps in edge-on projection. As the bulge gains angular
momentum, it starts picking up rotation. Signs of systematic rotation
are obvious at $t=2$~Gyr, and at $t=4$~Gyr, the ClB rotates faster,
with $(V/\sigma) \sim 0.5$ in the outer parts. It is interesting to
notice that the inner parts ($R < 2 R_{b,1/2}$) of this ClB do not
show much rotational motion - this mostly confined to the outer parts
of the ClB.  At the end of $4$~Gyr, the initially non-rotating ClB has
also become slightly rounder and hotter at the center.

\begin{figure}
\rotatebox{270}{\includegraphics[height=7.0 cm]{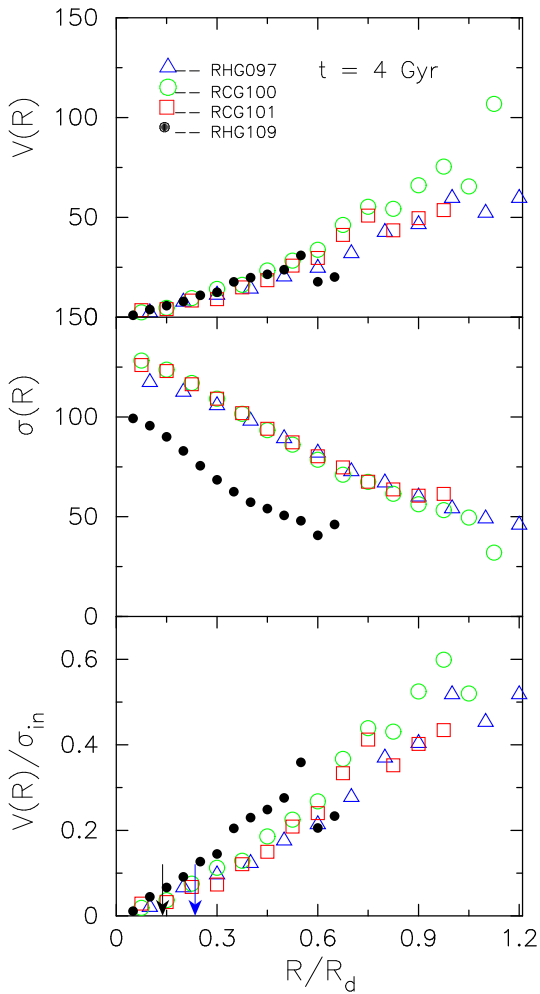}}
\caption{Rotation, velocity dispersion, and local $V/{\sigma}$ radial
  profiles for the four ClBs in models RHG097, RCG100, RCG101 and
  RHG109. The blue and black arrows show the half-mass radii for the
  two bulges of RHG109 and RHG097, respectively.}
\label{fig:vsig-radial}
\end{figure}

In Fig.~\ref{fig:vsig-radial}, we show radial profiles of rotational
velocity, $V(R)$, velocity dispersion, $\sigma(R)$, and local
$V(R)/\sigma_{\rm in}$, where $\sigma_{\rm in}$ denotes the average
velocity dispersion in the central region (within $0.5 R_{b,1/2}$),
for four ClBs after $4$~Gyr of evolution.  
We illustrate models RHG097, RCG100, RCG101 in order of decreasing bulge-to-disk 
mass ratio, and the relatively compact bulge model RHG109. The slope of the
rotational velocity is approximately the same for all four bulges
shown in the figure. This is also true for the remaining ClBs in
our sample. The mean velocity dispersion profiles are also nearly
identical for all these bulges except RHG109 which has lower velocity
dispersion throughout.  The $V/\sigma_{\rm in}$ profiles for the four
bulges are shown at the bottom panel of Fig.~\ref{fig:vsig-radial}. In
all cases, rotational motion becomes significant in the outer
parts of the bulge, for $R \ge 0.5 R_d \simeq 2 R_{b,{1/2}}$.

In some of the models in which the bar grows rapidly, reaching its
peak amplitude ($A_{2,max}$) within a billion years (e.g., RCG100,
RCG101), the bulge stars acquire significant rotation already within
$2$~Gyr. Thereafter, over the next $3 - 4$~Gyr, the rotational
profiles in these bulges change only insignificantly. On the other
hand, in models that grow their bar rather slowly, reaching peak
amplitude over $1 - 3$~Gyr, the rotation velocity in the ClBs
continues to change until the bar growth approximately saturates
at around $4$~Gyr.

Comparing the rotational profiles of our simulated ClBs with those in
observed barred galaxies is not straightforward because the kinematic
data in external galaxies include both the ClB stars and the stars in
the disk and a possible boxy bulge. Recently,
\cite{Fabriciusetal2012} analysed a large sample of galaxies
containing ClBs residing in both barred and non-barred galaxies. The
rotation profiles of the ClB host galaxies in their sample have a wide
range from shallow rising profiles, to steeply rising profiles often
associated with steeply falling velocity dispersion. Visual comparison
of the $V/\sigma$ profiles of their bulges with our simulated ClBs
suggest that some of the shallow-rising profiles could be similar to
our spun-up ClBs. The amplitude of the rotational profiles
($V/\sigma(R)$) for the ClB host galaxies reached within their bulge
radius is $\sim0.5-1.0$; see Fig.~13 of \cite{Fabriciusetal2012}. But
note that their bulge radius is not simply related to the effective
radius $R_e$; for their entire sample, it is typically $\sim 2R_e$.
The $V/\sigma(R)$ reached by the rotation profiles of our spun-up ClBs
are $\sim 0.2$ at $2 R_{b,1/2}$. Beyond this radius, $V/\sigma(R)$ for our ClBs
keeps increasing, reaching $V/\sigma (R)\sim 0.5$ at $\sim 4 R_{b,1/2}$.

\begin{figure}
\rotatebox{270}{\includegraphics[height=7.0 cm]{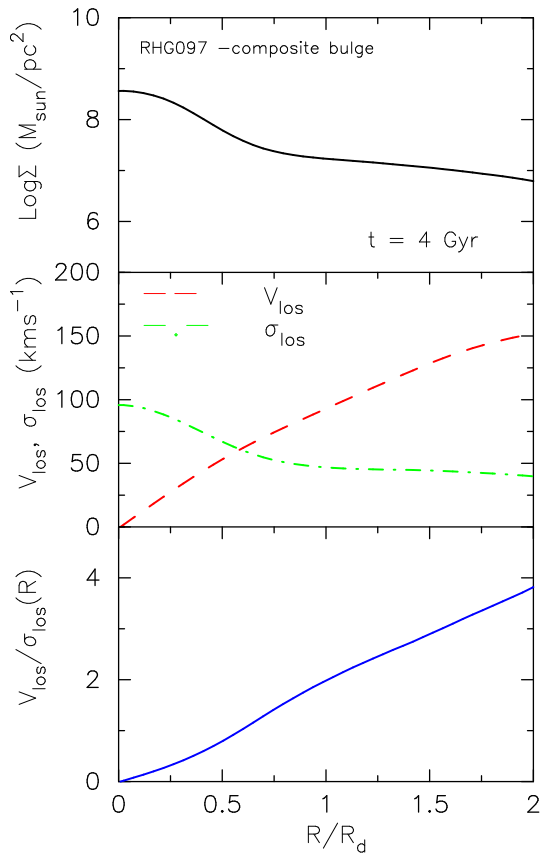}}
\caption{Surface density (top), mean line-of-sight velocity, and dispersion
  profiles (middle panel) for the composite bulge of model RHG097. The local
  $V/\sigma$ profile is shown in the bottom panel.}
\label{fig:denvelsigLOS97}
\end{figure}

However, the final bulge in our models is always a composite bulge
composed of the preexisting ClB and the boxy bulge formed from the
disk. Fig.~\ref{fig:denvelsigLOS97} shows the line-of-sight velocity,
dispersion and surface density profiles of the composite bulge (ClB
and boxy bulge combined) of model RHG097. The inner parts of this
galaxy model show a moderately rising rotation profile and declining
dispersion. The size of the boxy bulge is about $0.6 R_{\rm bar} \simeq
0.6 R_d$, not much greater than $\sim 2
R_{b,1/2}$ of the ClB, which is $\simeq 0.5 R_d$. At this radius, the local $V/\sigma \sim
1.0$. The slope of $V/\sigma (R)$ profile for this composite bulge
resembles many of the observed pseudobulges \citep{Fabriciusetal2012}. 
A more detailed comparison with observations is outside the scope of this paper.

Finally, we report here about the three-dimensional nature of the
induced rotational motion in our ClBs. None of the massive bulges in
our sample rotates cylindrically after $4 - 5$~Gyr, contrary to the
case of the low-mass ClB investigated by \cite{Sahaetal2012}.
Evidently, the orbital changes that occurred in the high-mass bulges
as a result of their interaction with the bar are weaker than in the
low-mass case.

\begin{figure}
\rotatebox{270}{\includegraphics[height=7.0 cm]{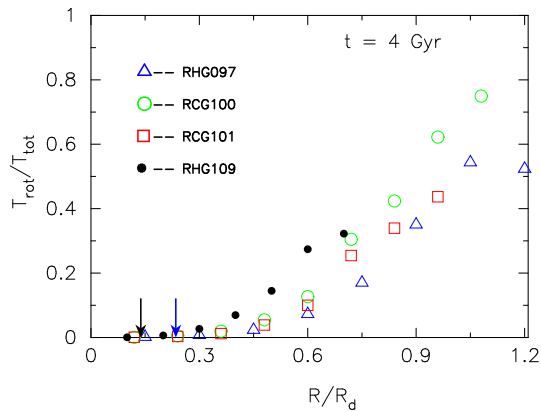}}
\caption{Fractional kinetic energy associated with the rotational
  motion for the four bulges in RHG097, RCG100, RCG101 and RHG109 in
  radial bins. The blue and black arrows show the half-mass radii for
  two bulges RHG109 and RHG097 respectively.}
\label{fig:Trot-radial}
\end{figure}

\begin{figure}
\rotatebox{270}{\includegraphics[height=7.0 cm]{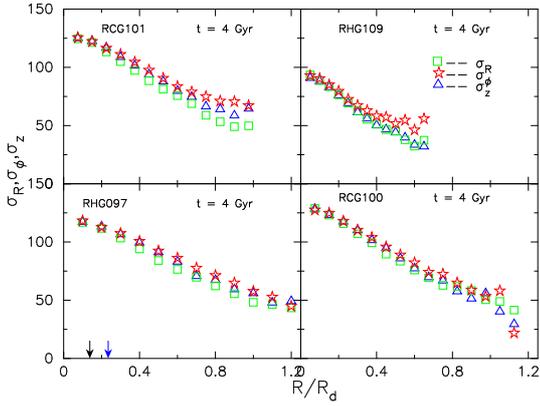}}
\caption{Radial variation of the radial, tangential, and vertical
  velocity dispersions in the four bulges in RHG097, RCG100, RCG101
  and RHG109 after 4 Gyr evolution. The blue and black arrows show the
  half-mass radii for two bulges RHG109 and RHG097, respectively.}
\label{fig:sig-radial}
\end{figure}

\subsection{Induced rotational kinetic energy and anisotropy}
\label{sec:aniso}

In Fig.~\ref{fig:vsig-radial}, we noticed that $V/\sigma_{\rm}$ rises
in the outer parts of the ClBs whereas the central parts of the bulge
remain with small rotation ($V/\sigma_{\rm} < 0.1$). This can also be
readily seen from the velocity maps of the ClB of RHG097 (see
Fig.~\ref{fig:2Dmap97}). Both these facts have prompted us to
investigate the actual rotational energy that is being imparted by the
bar to these bulges. The random kinetic energy of a group of stars
moving in the equatorial plane of the bulge, using cylindrical
coordinates, is given by

\begin{equation}
T_{{\rm rand}}(R_j)= T_{{\rm rand},R}(R_j) + T_{{\rm rand},\varphi}(R_j),
\end{equation}

\noindent where,

\begin{equation}
T_{{\rm rand},R}(R_j)= \frac{1}{2} \sum_{i=1}^{N_j} {{m_i [v_R(i)- {\bar{v}}_{R}]^2}}, 
\end{equation}
\noindent and

\begin{equation}
T_{{\rm rand},\varphi}(R_j)= \frac{1}{2} \sum_{i=1}^{N_j} {{m_i [v_{\varphi}(i) - {\bar{v}}_{\varphi}]^2}} 
\end{equation}

where $N_j$ is the number of bulge stars in the $j^{th}$ radial bin
and $R_j$ denotes its radius; $m_i$ is the mass of each bulge star,
$\bar{v}_R$ and $\bar{v}_{\varphi}$ are the mean radial and azimuthal
velocity of the stars.  

Similarly the rotational kinetic energy is given by
\begin{equation}
T_{\rm rot}(R_j) = \frac{1}{2} \sum_{i=1}^{N_j}{m_i {{\bar{v}_{\varphi}}^2}}
\end{equation}
so that the total kinetic energy of a group of stars moving in the
equatorial plane is
\begin{equation}
T_{\rm tot}(R_j) =  T_{\rm rand}(R_j) + T_{\rm rot}(R_j).
\end{equation}
Fig.~\ref{fig:Trot-radial} shows the radial distribution of the
fractional rotational kinetic energy in the final ClB bulges (used 
Eqs.~${1 - 5}$). The figure clearly shows that the inner regions
($R < 2 R_{b,{1/2}}$) of these ClBs do not acquire much rotation from
the bar, with final rotational kinetic energy amounting to about $5$\%
of the total planar kinetic energy after the evolution. It is the
outer regions which acquire most of the angular momentum transferred,
where for $R > 2 R_{b,{1/2}}$ typically $30 - 40$\% of the planar
kinetic energy is in rotation for most of the ClBs. Note a possible
explanation for this is given in section~\ref{sec:wakes}. In terms of
mass, this rotating region broadly contains about $40$\% or less of
the total bulge mass.

In Fig.~\ref{fig:sig-radial}, we show the radial variation of the
radial, azimuthal and vertical velocity dispersion ($\sigma_{R},
\sigma_{\varphi}$ and $\sigma_{z}$) for four ClBs at the end of
$4$~Gyr. Initially, all the bulges were isotropic by construction. Of
the four ClBs, RHG109 was spun-up significantly less than the other
three after $4$~Gyr of evolution. In all four ClBs, the final
azimuthal velocity dispersion is slightly higher compared to the
radial and vertical dispersions, with the largest effect in the
outskirts where rotation velocity is also highest. Thus
Fig.~\ref{fig:sig-radial} shows that these spun-up bulges tend to
become tangentially anisotropic.

\begin{figure}
\rotatebox{270}{\includegraphics[height=8.0 cm]{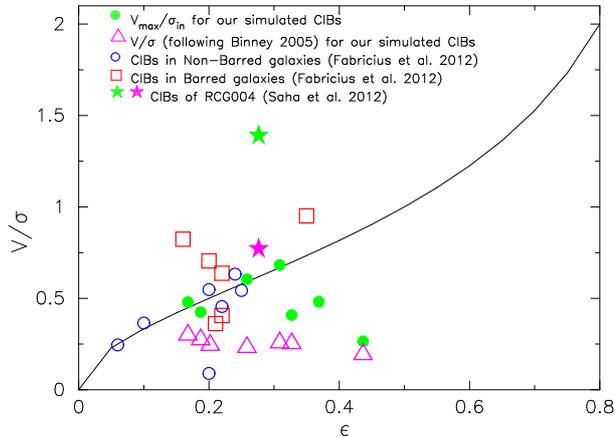}}
\caption{$V/{\sigma_{\rm}}$ versus ellipticity for all the ClBs in our
  galaxy models at the end of 4 Gyr. Initially all simulated ClBs were
  non-rotating. Over-plotted are the $V/{\sigma}$ values for ClBs from
  Fabricius et al. (2012).}
\label{fig:vsigeps}
\end{figure}

\subsection{$V/\sigma$ for the spun-up bulges }
\label{sec:spin}

Here we plot our spun-up ClBs on the ($V/{\sigma_{\rm}}, \epsilon$)
plot \citep{KI1982} to illustrate their rotational properties
further. Note that while this is easily done in simulations because of
the unique ID attached to each particle, comparing with observed ClBs
is non-trivial, because it is difficult to separate classical bulge
stars from stars in the boxy bulge in observations. First, we
calculate $V/{\sigma_{\rm}}$ values in the traditional sense. We
record the peak velocity, $V_m$, which is taken to be a simple average
of the outer velocities in the radial range $\sim 0.9 -1.2 R_d$
(excluding outliers such as the point with $\sim 100$km$s^{-1}$
in model RCG100, see Fig.~\ref{fig:vsig-radial}). This value is divided 
by the inner velocity dispersion, ${\sigma_{\rm}}_{\mathrm{in}}$,
computed by taking an average within the half-mass radius of the ClB
after $4$~Gyr. The resulting $V/{\sigma}$ values are listed as
$V_{\rm max}/{\sigma_{\rm}}_{\mathrm{in}}$ in Table~\ref{tab:paratab2}.

In addition, we also calculate $V/{\sigma_{\rm}}$ from the edge-on
kinematic maps of our evolved ClBs, following \cite{Binney2005}, and
denote these by $(V/{\sigma_{\rm}})_{<2\mathrm{d}>}$ . Thus, we
compute the surface density-weighted velocity and velocity dispersion
both in the inner ($R < 2 R_{b,1/2}$) and outer parts ($R > 2
R_{b,1/2}$). Consistent with the results of the last section, the
$(V/{\sigma_{\rm}})_{<2\mathrm{d}>}$ in the inner parts are small ($ <
0.1$). In the rest of the paper, we show the $(V/{\sigma_{\rm}})_{<2\mathrm{d}>}$ 
values for the outer parts only.

\begin{table}[b]
\caption[ ]{Bar size and bulge kinematics at the end of $4$~Gyr.}
\begin{flushleft}
\begin{tabular}{lcccccc}  \hline\hline
  Models    & $\frac{R_{{\rm bar}}}{R_d}$ & $\frac{R_b}{R_{{\rm bar}}}$ &$\epsilon$ & $V_{\rm max}/\sigma_{\mathrm{in}}$  & $(V/{\sigma_{\rm}})_{<2\mathrm{d}>}$ \\
             &  &  & & & ($R>2R_{b,1/2}$)\\
\hline
\hline
RCG076     &0.61 &3.42 & 0.37 & 0.48 & 0.29\\
RCG110     &0.83 &0.97 & 0.33 & 0.41 & 0.25\\
RHG109     &0.90 &0.72 & 0.44 & 0.26 & 0.19\\
RCG101     &1.10 &0.96 & 0.31 & 0.68 & 0.26\\
RCG100     &1.03 &1.04 & 0.29 & 0.60 & 0.23\\
RHG099     &0.94 &1.15 & 0.19 & 0.42 & 0.27\\
RHG097     &0.93 &1.31 & 0.17 & 0.48 & 0.30\\
RCG004	   &0.85 &1.52 & 0.28 & 1.40  & 0.77\\
\hline
\end{tabular}
\end{flushleft}
\label{tab:paratab2}
\end{table}

We calculate the ellipticities ($\epsilon$) of the ClBs by the method of 
diagonalizing the moment-of-inertia tensor. The intrinsic ellipticity of 
the bulge is then $\epsilon = 1- c/a$, where $c/a$ is the ratio of the minor 
to major axis. In edge-on projection (where the inclination angle is $90^o$ w.r.t. the
LOS), the apparent ellipticity would be equal to the intrinsic
ellipticity derived using the moment-of-inertia method. We compared
these axis ratios with those obtained using IRAF ellipse fitting, and
the agreement between both methods is fairly good.
 
Fig.~\ref{fig:vsigeps} shows the well-known $V/{\sigma_{\rm}} -
\epsilon$ plot for all the ClBs in our model galaxies after $4$~Gyr.
Ellipticities and $V/{\sigma_{\rm}}$ values at this time are also
tabulated in Table~\ref{tab:paratab2}. $4$ of the initially
non-rotating ClBs from our simulated sample have final $V_{\rm
  max}/{\sigma_{\rm}}_{\mathrm{in}}$ values which formally put them on
the oblate isotropic rotator line \citep{Binney1978}, while the
remaining $3$ bulges did not spin up as much. For all our ClBs, the
outer $(V/{\sigma_{\rm}})_{<2\mathrm{d}>}$ values are lower
compared to the $V_{\rm max}/{\sigma_{\rm}}_{\mathrm{in}}$. This is
because the induced rotation profiles in these ClBs have a shallow
rise and reach their maxima in the outermost parts which contain
little mass. The final $(V/{\sigma_{\rm}})_{<2\mathrm{d}>}$ may be
slightly on the low side in our models, however, because the initial
ClBs were modelled as King profiles in which the density of stars drops to zero
rapidly in the outer parts marked by the truncation radii and contain fewer stars.
In contrast, the observed ClBs which follow Sersic profiles $r^{1/n}$ \citep[e.g.,][]{Caonetal1993}, 
with $n \sim 4$ have comparatively less steeply falling density and contains larger
number of stars at the outer parts. Our analysis reveals that it is the outer
part of the massive bulges which gain most of the angular momentum transferred by the
bar. Having these outer parts well populated by stars on stable orbits would help
resonant transfer of angular momentum more effectively or vice-versa. We, therefore, 
expect angular momentum transfer and the bulge spin up to be greater for less steep bulge
profiles as might be the case in observed ClBs.

The two star symbols show the equivalent $V/{\sigma_{\rm}}$ values for
the low mass ClB model of RCG004 from \citet{Sahaetal2012}. They are
considerably higher than for the massive bulge models reported
here. The ClB in this model had significantly lower mass ($0.067M_d$)
but similar $R_{b,1/2}/R_d\simeq 0.21$. Clarifying the origin of this
difference will require a more extensive parameter study of the
evolution of low mass bulges in bar unstable disks.

\begin{figure}
\rotatebox{270}{\includegraphics[height=8.0 cm]{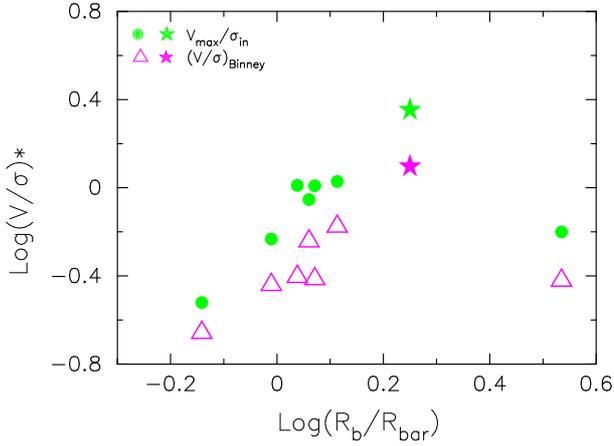}}
\caption{Dependence of $V/\sigma$ on the ratio of bulge size to bar
  size of the embedded ClBs at the end of 4 Gyr. Initially, all the
  bulges were non-rotating. The star signs indicate the ClB of RCG004,
  as in Fig.~\ref{fig:vsigeps}.}
\label{fig:vsigRbar}
\end{figure}

In addition to our simulated ClBs, we have overplotted in
Fig.~\ref{fig:vsigeps}, $6$ ClBs in barred galaxies (NGC1023,
NGC2859, NGC2880, NGC3521, NGC3992 and NGC4260) and $7$ ClBs in
non-barred galaxies (NGC4698, NGC4772, NGC3898, NGC3245, NGC3031,
NGC2775, NGC2841) taken from \cite{Fabriciusetal2012}. In most cases,
the ClBs in barred galaxies rotate faster than their counterparts in
non-barred galaxies. Whether in these galaxies the bar might be
responsible, in part, for spinning up these ClBs will require a
separate investigation.

Fig.~\ref{fig:vsigRbar} shows the dependence of normalized
$(V/{\sigma_{\rm}})^*$ on the ratio of bulge size to bar size
($R_b/R_{\rm bar}$). For the normalization, we used the relation
$V/{\sigma_{\rm}} \sim \sqrt{\epsilon/(1 -\epsilon)}$. For all bulges
(except RCG076) the final $(V/{\sigma_{\rm}})^*$ at $4$~Gyr, as well
as $(V/{\sigma_{\rm}})_{<2\mathrm{d}>}$ following \cite{Binney2005}
correlate tightly with increasing ratio $R_b/R_{\rm bar}$.  This
suggests that the dynamics of the ClB spin-up depends on the relative
size of the bar. It is possible that there is an optimal size for the
rotating bar to provide the most angular momentum transfer to the ClB
through resonant angular momentum transfer and dynamical friction on
the bulge stars (see Section~\ref{sec:wakes}); however, this needs
additional models with large ClBs.  Certainly, from the models studied
here, the spin-up of ClBs is not efficient when the ClB is too
compact.

\section{Conclusions and discussion}
\label{sec:discus}
We have used self-consistent N-body simulations of disk galaxies to
study the interaction of massive ClBs with the bar that forms in the
disk. We found that significant angular momentum is transferred by
resonant gravitational interaction from the bar to the ClB in all cases, in a
time-scale of a few Gyr. We have analyzed the angular momentum
transfer using orbital frequency analysis and examining the barred
density wake in the bulge. Because in our models most of the angular momentum
is transferred to the outer parts of the ClB, it is likely that the
induced rotation in Sersic bulges would be higher than in those studied
here. Our main conclusions from this work are as follows:

\noindent 1. Massive classical bulges gain as much specific angular
momentum through spin-up by the bar as do lower-mass bulges.

\noindent 2. Most of the angular momentum transfer occurs through
low-order resonances.  In particular, we found that a lot of angular
momentum can be transferred via $5:2$ resonant orbits which are common
in the models studied here.

\noindent 3. The bar generates a density wake in the bulge due to
trapped 2:1 orbits. For lower mass bulges in our sample, such a wake
is found to be misaligned with the bar and causes angular momentum
transfer while for higher mass bulges the misalignment is not
substantial. At later phases of evolution, the bulge density wake
aligns with the bar and evolves into the bulge-bar.

\noindent 4. The spin-up process creates a characteristic linear
rotation profile such that significant rotation is induced beyond
$\sim2 R_{b,1/2}$. Typical mass-weighted $V/\sigma$ beyond $\sim 2
R_{b,1/2}$ are $\sim 0.2$, and the local $V_{\rm max}/\sigma_{\mathrm{in}}$ 
reached at the largest radii are $\sim 0.5$.

\noindent 5. Contrary to the case of the low-mass ClBs studied in
\citet{Sahaetal2012, SahaGerhard2013a}, the rotation induced in the ClB
is not cylindrical.

\noindent 6. The spin-up process also creates mild tangential anisotropy
in the outer regions of the ClBs studied here.

\noindent 7. In all models a box/peanut bulge also forms through the
bar-buckling instability. The final systems therefore have a composite bulge
with the ClB superposed on the slightly larger box/peanut bulge. This
suggests that composite bulges may be common in galaxies.

The main result of this paper is that also comparatively massive
bulges are impacted by the angular momentum transfer mechanism from
the bar, which was previously investigated only for low-mass ClBs,
both non-rotating \citep{Sahaetal2012} and rotating
\citep{SahaGerhard2013a}. In our sequence of models, the spin-up is
most efficient when the size of the bar is somewhat larger than the
ClB, but not too large. This suggests that larger effects might be
expected in the presence of gas, which would favour higher pattern
speeds and mass inflow. If nuclear bars formed as a result of central
mass build-up, they could lead to additional angular momentum
transfer. These effects merit future investigation.

\section*{Acknowledgement}
The authors thank the referee, Frederic Bournaud, for thoughtful comments which helped
improving the clarity of the paper.

\end{document}